\documentclass[aps,prb,twocolumn,showpacs,floats,epsfig,superscriptaddress]{revtex4}
\usepackage{amsmath}
\usepackage{graphicx}
\usepackage{graphics}
\usepackage{amssymb}
\usepackage{epsfig}
\usepackage{times}

\begin{document}

\title{Biaxial spin-nematic phase of two dimensional disordered rotor models 
and spin-one bosons in optical lattices}

\author{Jean-S\'ebastien Bernier}
\affiliation{Department of Physics, University of Toronto, Toronto,
Ontario, Canada M5S 1A7}

\author{K. Sengupta}
\affiliation{Theoretical Condensed Matter Physics Division, Saha Institute of 
Nuclear Physics, 1/AF Bidhannagar, Kolkata-700064, India}

\author{Yong Baek Kim}
\affiliation{Department of Physics, University of Toronto, Toronto,
Ontario, Canada M5S 1A7}
\affiliation{School of Physics, Korea Institute for Advanced Study, Seoul 130-722,
Korea}

\date{\today}

\begin{abstract}
We show that the ground state of disordered rotor models with
quadrupolar interactions can exhibit biaxial nematic ordering in the
disorder-averaged sense. We present a mean-field analysis of the
model and demonstrate that the biaxial phase is stable against small
quantum fluctuations. We point out the possibility of experimental
realization of such rotor models using ultracold spin-one Bose atoms
in a spin-dependent and disordered optical lattice in the limit of a
large number of atoms per site and also suggest an imaging
experiment to detect the biaxial nematicity in such systems.
\end{abstract}


\maketitle

\section{Introduction}

Nematic phases of matter have been widely studied in many different
areas of condensed matter physics \cite{dg}. For example, consider a
system of molecules having some vector quantity (say a polarization
vector ${\bf p}$) associated with each of them. Since molecules are
extended objects, a local coordinate system can be attached to each
molecule. Hence, for a molecule at a point
$i$, one can define a local nematic order parameter: $Q^{\alpha
\beta}_i = p_{i \alpha} p_{i \beta} -\sum_{\alpha} p_{i \alpha}^2/3$
\cite{dg}. The nematic phase is then defined as the state of the
system for which $\sum_{i} {\bf p}_i =0$ whereas $\sum_i Q^{\alpha
\beta}_i$ becomes a traceless $3\times 3$ matrix with eigenvalues
$Q= \left[-(Q_1+Q_2),Q_1,Q_2\right]$. Such phases are of two types:
uniaxial nematic where $Q_1=Q_2$ and biaxial nematic for which $Q_1
\ne Q_2$\cite{dg}. In both these phases, the system breaks
rotational symmetries while preserving translational invariance.
Uniaxial nematic phases occurs more frequently whereas the first
experimental observation of biaxial nematic phase in a classical
system of molecules has only been
recently reported \cite{merkel}. Although much more difficult to
realize, biaxial nematic phases are of great theoretical interest
since they are known to support non-Abelian defects \cite{Mermin}.
Uniaxial nematic phases have also been studied in the context of several
quantum condensed matter systems such as strongly
correlated electron systems \cite{halboth,oganesyan,ybk,kivelson,
andreev,chandra,gorkov} and ultracold spin-one atoms
in optical lattice\cite{Demler1,Adi1,snoek1}. For example, in the case 
of spin-one boson system, in some parameter regime, the ground state breaks
spin-rotational invariance along one of the axes while maintaining
translational invariance with no spin-order $\left<{\bf S}\right> =0
$ and with two equal non-zero eigenvalues of the matrix $\left
<S^{\alpha}S^{\beta} \right>$. However, there has been no proposals
of realization of biaxial nematic ground states in the context of
quantum condensed matter systems using standard local Hamiltonians.

In this work, we point out that disordered ${\rm O}(2)$ rotor models
with quadrupolar interaction in two dimensions (2D) can exhibit
biaxial nematic phase in the disordered averaged sense. Such rotor
models are described by the Hamiltonian
\begin{eqnarray}
{\mathcal H}_{\rm rotor} &=& U \sum_i L_i^2 - g \sum_{<ij>}
\left(\sum_{a=x,y,z} d_{ia} \Lambda_{ij}^a  d_{ja}  \right)^2
\label{model1}
\end{eqnarray}
where ${\bf d} \equiv (d^x,d^y,d^z) = \left(\sin\theta\cos\phi,
\sin\theta\sin\phi,\cos\theta \right)$ is a unit vector expressed
in terms of the rotor angles $\theta$ and $\phi$, $L_i$ denotes
angular momentum of the rotor, $\sum_{\left<ij\right>}$ denotes sum
over nearest neighbor sites, and ${\bf \Lambda}_{ij}$ are
dimensionless random couplings between the unit vectors ${\bf d}$ at
sites $i$ and $j$. In the following we shall consider $\Lambda^x =
\Lambda^y \ne \Lambda^z$, and choose $\Lambda^{x(z)}$ to have
Gaussian distributions with same width $\delta \Lambda$ and means
${\bar \Lambda}^{x(z)}$. The main result of this work is that in the
limit $\left|{\bar \Lambda}^{x}-{\bar
\Lambda}^{z}\right|/ \delta \Lambda \ll 1$, the model given by Eq.\ \ref{model1}
exhibits a biaxial nematic ground state in the disordered average
sense for $U \ll g$. We also show, using a mean-field analysis, that
the biaxial phase is stable against quantum fluctuations until
$\nu=g/U$ is reduced to a critical value $\nu_c$. Note that the
above-mentioned regime can also be reached for arbitrarily weak
disorder: $\delta \Lambda \ll {\bar \Lambda}^x, {\bar \Lambda}^z$
provided that $\left|{\bar \Lambda}^{x}-{\bar \Lambda}^{z}\right| \ll 
\delta \Lambda$. This particular feature of the rotor model, as we 
shall see, makes it easily realizable using ultracold atoms in optical lattices.

The choice of the rotor Hamiltonian (Eq.\ \ref{model1}) is not
arbitrary, but is motivated by their connection with ultracold
spin-one bosons in an optical lattice. The low energy effective
Hamiltonian of such bosons systems has been derived in the absence
of disorder in Refs.\ \onlinecite{Demler1,Adi1,snoek1}. It has been
shown that the effective theory of the Mott phases of such a system
can be described, in the limit of large number of bosons per site,
by rotor models similar to Eq.\ \ref{model1}, supplemented by an
additional constraint on the number $N$ and the total spin $S$ of
the bosons per site: $N+S = {\rm even}$
\cite{Demler1,Adi1,snoek1}. We generalize this analysis to include
disorder and show that in the presence of spin-dependent disordered
optical lattices, which can be easily created by using a speckled
laser field \cite{horak,lye}, the Mott phases of ultracold spin-one
bosons is indeed described by Eq.\ \ref{model1} supplemented by the
constraint condition mentioned above. Further, as will be shown
later, for $U \ll g$, the constraint condition becomes irrelevant
and in this limit we expect our analysis of the rotor model to apply
for the bosons as well. This correspondence therefore allows us to
suggest experiments on ultracold atoms which can in principle probe
such a biaxial nematic phase. We suggest a straightforward imaging
of ultracold atoms using a polarized laser beam which can
distinguish the biaxial phase from other spin or uniaxial nematic
phases.

The organization of the paper is as follows. In Sec.\ \ref{rotor},
we analyze the rotor model (Eq.\ \ref{model1}) and demonstrate
that it exhibits a biaxial nematic ground state. This is followed by
Sec.\ \ref{effective}, where we derive the effective low-energy
Hamiltonian of spin-one ultracold Bosons in a spin-dependent
disordered optical lattice and discuss its relation to the rotor
model analyzed in Sec.\ \ref{rotor}. In Sec.\ref{detection}, we
suggest an experiment to detect biaxial nematic order. Finally, we
conclude in Sec.\ref{conclu}.

\section{Analysis of the Rotor Model\label{rotor}}

In this section, we first present an analysis of the rotor model
(Eq.\ \ref{model1}) and demonstrate the presence of biaxial nematicity
at $U = 0$. This is followed by the derivation and analysis of a mean
field theory for finite $U$ which illustrates  the stability of the
biaxial phase against small quantum fluctuations.

\subsection{Limit of $U=0$}

For $U=0$, in the absence of quantum fluctuations, the Hamiltonian
(Eq,\ \ref{model1}) is diagonal in the $|\theta,\phi\rangle$ basis.
This allows us to write
\begin{equation}
H'_{\rm rotor}  = -g \sum_{\left<ij\right>}A^2_{ij}, \label{hamu0}
\end{equation}
where $A_{ij}$ is given by
\begin{eqnarray}
A_{ij} = \Lambda^x_{ij} \sin\theta_i \sin\theta_j
\cos(\phi_i-\phi_j) + \Lambda^z_{ij} \cos\theta_i \cos\theta_j
\label{aij}
\end{eqnarray}
with $0 \le \theta_i \le \pi$ and $0\le \phi_i < 2\pi$. Let us now
consider a ground state configuration of the Hamiltonian $H'_{\rm
rotor}$ given by
\begin{eqnarray}
\Psi_G = \left|\theta_1^0,\phi_1^0; \theta_2^0,\phi_2^0; ...
\theta_i^0,\phi_i^0; .... \theta_N^0,\phi_N^0; ....\right>
\label{groundwave1}
\end{eqnarray}
where $(\theta_i^0,\phi_i^0)$ denotes the value of the angles
$\theta$ and $\phi$ at the $i$th site in the ground state. From
Eqs.\ \ref{hamu0} and  \ref{aij}, one can immediately see that
$H'_{\rm rotor}$ has an infinite number of degenerate ground states
since any local transformation $\theta_i^0 \rightarrow \pi -
\theta_i^0$, $\phi_i^0 \rightarrow \phi_i^0 + \pi$ leaves $H'_{\rm
rotor}$ invariant for any random set of $\Lambda^{x,z}_{ij}$. It is
also easy to see that such a local transformation changes ${\bf d}_i
\rightarrow -{\bf d}_i$ and leaves the local nematic order parameter
$ Q^{ab}_i = d_i^a d_i^b - \delta_{ab}/3$ invariant. Hence for the
purpose of computing the nematic order parameter, we can choose any
one of these ground state configurations.

The rotor Hamiltonian $H'_{\rm rot}$ is also invariant under two
global transformations. The first of them  $T_1: \phi \rightarrow
\phi + \eta $ is a gauge freedom which allows us to choose the
orientation of the global $x$ axis. The second transformation
$T_2:\theta \rightarrow \pi-\theta$ leaves the diagonal components
of $Q^i_{ab}$ invariant while changing the off-diagonal components.
The ground states $\Psi_G$ (Eq.\ \ref{groundwave1}) and $T_2 \Psi_G
= \Psi'_G$ are degenerate.

Using the above-mentioned local and global transformation
properties, we can considerably simplify the numerical procedure for
obtaining ground states of $H'_{\rm rotor}$ for a given disorder
realization. First, from Eq.\ \ref{aij}, we find that any
ground-state configuration $\Psi_G$ will always have $\phi_i-\phi_j
=0,\pi$. Using the transformation $T_1$ , we can therefore choose
each individual $\phi_i$ to be either $0$ or $\pi$. This is a gauge
choice and, hence, does not affect any physical quantity. Next instead of
using $\Psi_G$ to compute $Q_{ab}^i$, we find the state
$\Psi^{''}_G$, for which $\phi_i =0$ at every site. This can be done
since $\Psi^{''}_G$ is always a part of the degenerate states which
can be reached from $\Psi_G$ by successive application of the
above-mentioned local transformation. Using these two facts, we see
that for any set of random coefficients $\Lambda^{x,z}_{ij}$, one
can compute $Q_{ab} = \sum_{i}d_{ia} d_{ib} - \delta_{ab} /3$ by
finding the ground state configuration of $H^{''}_{\rm rotor} = -g
\sum_{<ij>} (A'_{ij})^{2}$ with
\begin{eqnarray}
A'_{ij} = \Lambda^x_{ij} \sin\theta_i \sin\theta_j + \Lambda^z_{ij}
\cos\theta_i \cos\theta_j.
\end{eqnarray}
The ground states of $H^{''}_{\rm rotor}$ is two-fold degenerate and
these two ground states $\Psi^{(1)}_G$ and $\Psi_G^{(2)} = T_2
\Psi^{(1)}_G$ are related by the global transformation $T_2$. The
two ground states have opposite signs for off-diagonal elements of
$Q_{ab}$ in the chosen gauge and therefore after averaging over
these ground states, the off-diagonal elements of $Q_{ab}$ vanishes.
Thus we are finally left with the diagonal components $ Q_{aa} = 
(-(Q_1+Q_2),Q_1,Q_2) =  (C-1/3, -1/3, -(C-2/3))$, 
where $0 \leq C \leq 1$,  
for a given set of random coefficients
$\Lambda^{x,z}_{ij}$. Finally, one can numerically average over
different realizations ${\bar Q}_{a}= \left<Q_{aa}\right>_{\rm
disorder}$ to obtain the disorder-averaged nematic order parameter.
The biaxial nematic phase is realized when ${\bar C} \ne 0$, 
$\frac{1}{2}$ or $1$.

The ground state of $H{''}_{\rm rotor}$ can be obtained numerically
using standard minimization procedure for each disorder realization.
For the sake of brevity, we choose $\Lambda_{ij}^{x(z)}$ to be a set
of random numbers having a Gaussian distribution with same standard
deviation $\delta \Lambda$ and different means ${\bar \Lambda}^{x(z)}$ 
and average over $500$ disorder realizations. We choose ${\bar
\Lambda}^x > {\bar \Lambda^z}$ here without loss of generality. We
study finite-size systems with sizes ranging from $N_S= 8
\times 8$ to $N_S= 20 \times 20$ with periodic boundary conditions and
use standard $1/N_S$ extrapolation to obtain the infinite system size
limit\cite{comment2}. A few comments about the numerical procedure
are in order here.

First, for $({\bar\Lambda}^{x}-{\bar \Lambda}^{z})
/\delta \Lambda \gg 1$, it is clear that the system energy reaches its minimum
at $\sin\theta_i \sin\theta_j = 1$ and $\cos\theta_i \cos\theta_j= 0$
for all sites and disorder realizations. Consequently, after
disorder-averaging, ${\bar C}=1$, and the ground state is an
uniaxial nematic phase.

Second, in the limit $({\bar \Lambda}^{x}-{\bar
\Lambda}^{z})/\delta \Lambda \ll 1$, for a given disorder realization, we find that
the configuration with minimum energy is part of a group of three
possible solutions. The first corresponds to
$\sin\theta_i\sin\theta_j = 1$ and $\cos\theta_i\cos\theta_j= 0$ for
all sites. For a single disorder realization with this solution, the
order parameter then has $C=1$. The second solution possibly
assumed by the system is $\sin\theta_i\sin\theta_j = 0$ and
$\cos\theta_i\cos\theta_j = \pm 1$ for all sites. The value of the
order parameter corresponding to such a disorder realization is
then given by $C = 0$. Finally, the third possible solution
corresponds to a configuration for which $\sin\theta_i\sin\theta_j$
and $\cos\theta_i\cos\theta_j$ are neither $0$ nor $\pm 1$, and are
varying in values from site to site. For such a disorder
realization, $C \ne 0$, $\frac{1}{2}$ or $1$ and one has a biaxial nematic. 
However, having $C \ne 0$, $\frac{1}{2}$ or $1$ for a single disorder 
realization does not guarantee that the system is biaxially ordered. 
The system can in fact be separated into domains, each domain representing one 
of the three possible solutions. For example, the system could be separated 
into domains representing only the first and second type of solutions and still have 
$C \ne 0$, $\frac{1}{2}$ or $1$. Nonetheless, this situation is unlikely;
in most cases where $C \ne 0$, $\frac{1}{2}$ or $1$ we find that the system adopts 
the third solution. Nevertheless, as mentioned previously, to find the physically 
meaningful ground state, we need to average over many disorder realizations 
and by doing so restore the translational invariance of the system. Consequently, 
${\bar C} \ne 0$, $\frac{1}{2}$ or $1$ can be used only after disorder-averaging
to determine whether the system is biaxially ordered for certain disorder strength.

\begin{figure}
\rotatebox{270}{
\includegraphics[width=6cm]{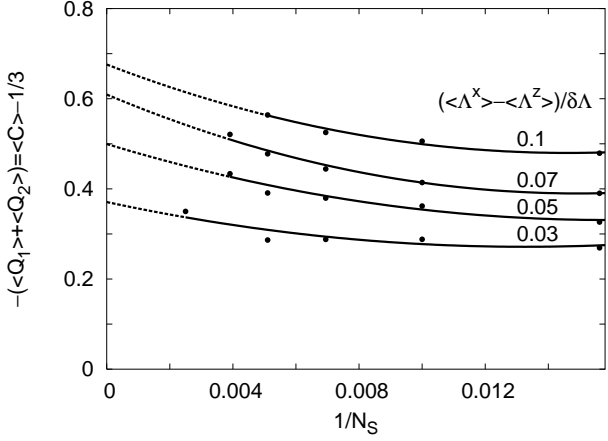}}
\caption{$-({\bar Q}_1+{\bar Q}_2) = {\bar C}-1/3$ as a function of the inverse lattice
size for several disorder strengths. For moderate disorder 
$({\bar\Lambda}^x-{\bar \Lambda}^z)/\delta \Lambda \ge 0.1$, 
an extrapolation
to large system size shows that the ground-state will be uniaxially
ordered, ${\bar C} \rightarrow 1$. For strong disorder 
$({\bar \Lambda}-{\bar \Lambda}^z)/\delta \Lambda \le 0.07$, the ground-state
remains biaxially ordered, ${\bar C} \ne 1$ after disorder
averaging within $1/N_S$ extrapolation. [Note: ${\bar Q}_1 = \langle Q_1 \rangle$,
${\bar Q}_2 = \langle Q_2 \rangle$, ${\bar C} = \langle C \rangle$,
${\bar \Lambda}^x = \langle \Lambda^x \rangle$, 
${\bar \Lambda}^z = \langle \Lambda^z \rangle$]} \label{fig1}
\end{figure}

The result of the numerical calculation, which corroborates the
above discussion, is shown in Fig.\ \ref{fig1}. 
After fitting second order polynomial functions to our data and
extrapolating the resulting functions to $1/N_S \rightarrow 0$,
we find that for
$({\bar \Lambda}^x-{\bar \Lambda}^z)/\delta \Lambda \le 0.07$, the
ground state is biaxial nematic with ${\bar C} \ne 1$. For
intermediate or larger values of $({\bar \Lambda}^x-{\bar
\Lambda}^z)/ \delta \Lambda \ge 0.1$, ${\bar C}=1$ and we have an
uniaxial nematic ground state. So this demonstrates that for $U=0$,
the rotor model exhibits a biaxial nematic ground state in the
disordered average sense in the limit $({\bar\Lambda}^x-
{\bar \Lambda}^z)/\delta \Lambda \ll 1$. In the next subsection, we
study the effect of quantum fluctuations on this state.

A comment about the validity of the $1/N_S$ extrapolation is in order
here. This procedure definitely leaves open a possibility that a
large enough system size has not been reached in our numerics and
the biaxial nematic state seen here is an artifact of finite system
size. The resolution of this question, within our numerical
procedure, is not straightforward. However, we can be certain that
even if this is the case, with small enough $({\bar\Lambda}^x
-{\bar \Lambda}^z)/\delta \Lambda$, the biaxial phase will manifest
itself over a large window of energy scale at finite temperature,
even if the zero temperature ground state turns out to be uniaxial
nematic.

\subsection{Quantum fluctuations for $U \ne 0$}
\label{effac}

To study the effect of quantum fluctuations, it is more convenient
to switch to a path integral representation of the rotor Hamiltonian
(Eq.\ \ref{model1}). To this end, we first write the partition
function of our Hamiltonian as
\begin{eqnarray}
Z &=& {\rm Tr} \exp(-\beta {\mathcal H}_{\rm rotor}) = {\rm Tr}
\exp\left[-\beta(T+V)\right]
\label{pf1} \\
T &=& U \sum_i L_i^2 \label {kin1}\\
V &=& -g \sum_{abcd} \sum_{\langle ij\rangle} d_{ia} d_{jb}
\Gamma^{abcd}_{ij} d_{ic} d_{jd} \label{potential}
\end{eqnarray}
where we have introduced the notation $\Gamma^{abcd}_{ij} =
\Lambda^a_{ij} \Lambda^c_{ij} \delta_{ab} \delta_{cd}$. To obtain
the partition function we calculate the trace in (\ref{pf1}) by
writing it as a path integral over $M$ time slices, and then
inserting a complete set of coherent states $|\theta,\phi\rangle$
over each slice, so that
\begin{eqnarray}
Z &=& \lim_{M\rightarrow \infty} {\rm Tr}
\left[\exp\left\{-\delta \tau (T+V)\right \}\right]^M  \nonumber\\
&=& \int {\mathcal D}\theta {\mathcal D}\phi \prod_{\alpha=0}^{M-1}
\left< \theta(\tau_{\alpha+1});\phi(\tau_{\alpha+1})\right|
\exp\left[-\delta \tau T\right] \nonumber\\
&& \times \exp\left[-\delta\tau V\right]
\left|\theta(\tau_{\alpha}); \phi(\tau_{\alpha})\right> \label{pf2}
\end{eqnarray}
where $\delta \tau = \beta/M$.
Inserting a set of complete states $|l,m\rangle$ (defined by $\langle\theta,\phi|l,m\rangle =
Y_{lm}(\theta,\phi)$, where $Y_{lm}$ are the spherical harmonics) on each
slice, we get
\begin{eqnarray}
Z &=&  \int {\mathcal D}\theta {\mathcal D}\phi \prod_{\alpha=0}^{M-1}
\exp\left[-\delta \tau V(\theta,\phi)\right] T_{\alpha} \label{pf3} \\
T_{\alpha} &=& \sum_{l_{\alpha}=0}^{\infty}
\sum_{m_{\alpha}=-l_{\alpha}}^{l_{\alpha}} \prod_{i} \exp\left[-U \delta\tau
l_{\alpha}(i)\left(l_{\alpha}(i)+1\right)\right] \nonumber\\
&& \times Y_{l_{\alpha}(i),m_{\alpha}(i)}^{\ast}
\left(\theta_{\alpha+1}(i),\phi_{\alpha+1}(i) \right) \nonumber\\
&& \times Y_{l_{\alpha}(i),m_{\alpha}(i)}
\left(\theta_{\alpha}(i),\phi_{\alpha}(i) \right) \label{kin2}
\end{eqnarray}
To rewrite (\ref{kin2}) in a suitable form, we use the identities \cite{kleinert}
\begin{eqnarray}
&& e^{h \cos\left(\Delta \theta \right)} = \sqrt{\frac{8\pi^3}{h}}
\sum_{l=0}^{\infty} \sum_{m=-l}^{l} I_{l+\frac{1}{2}}(h)
Y_{lm}^{\ast}(\theta,\phi) Y_{lm}(\theta',\phi') \nonumber\\
&& \lim_{h\rightarrow \infty} I_{l+\frac{1}{2}}(h) =
\exp\left[-\frac{\left(l+1/2\right)^2-1/4}{2h}\right] + O(1/h^2) \nonumber \\
\end{eqnarray}
where $I_{l+\frac{1}{2}}$ is the modified Bessel function and $\cos\Delta\theta=
\cos\theta\cos\theta'+\sin\theta\sin\theta'\cos(\phi-\phi')$.
Then, after a few straightforward
manipulations, we get in terms of ${\bf d}=(\sin\theta\cos\phi,\sin\theta\sin\phi,
\cos\theta$) fields,
\begin{eqnarray}
Z &=& \int {\mathcal D}{\bf d} \prod_{i} \delta
\left(\sum_{a}|d_{ia}|^2-1\right) \exp\left(-S\right) \label{pf4} \\
S &=& \int d\tau \left[ \sum_{ia} \left(\partial_{\tau}d_{ia} \right)^2 -
\nu \sum_{<ij>} \sum_{abcd} d_{ia} d_{ic} \Gamma^{abcd}_{ij} d_{jb} d_{jd} \right] \nonumber\\
\label{efact1}
\end{eqnarray}
where we have rescaled $\delta\tau$ so that $4U\delta\tau=1$ and $\nu
= g/4U$.

Next, we introduce a Lagrange multiplier field $\lambda_i(\tau)$ to
implement the constraint and decouple the quartic term in $S$ using
an auxiliary field $N_{ab}^i$. After some algebra, we get
\begin{eqnarray}
Z &=& \int {\mathcal D}{\bf d} {\mathcal D} N {\mathcal D }\lambda
\exp\left(-S_1 \right) \label{pf5} \\
S_{1} &=& \int d\tau \Bigg[ \sum_{i} \Bigg [ \sum_a \left(\partial_{\tau}d_{ia} \right)^2
+i\lambda_i \left(\sum_a |d_{ia}|^2 -1 \right) \nonumber\\
&& - i \sum_{ab} d_{ia}d_{ib} N_{ab}^i \Bigg]
- \frac{1}{2\nu} \sum_{<ij>}
\sum_{abcd} N_{ac}^i (\Gamma^{abcd}_{ij})^{-1} N_{bd}^j \Bigg] \nonumber\\
\label{mfac1}
\end{eqnarray}
where the auxiliary fields $N_{ab}^i$ are not the order parameter
fields, but their conjugate. The order parameter fields $P_{ab}^i$
can now be introduced by a second Hubbard-Stratonovitch
transformation which yields
\begin{eqnarray}
Z &=& \int {\mathcal D}{\bf d} {\mathcal D} N {\mathcal D }\lambda {\mathcal D}P
\exp\left(-S_{\rm eff} \right) \label{pf6} \\
S_{\rm eff} &=& \int d\tau \Bigg[ \sum_{i} \Bigg \{ \sum_a \left(\partial_{\tau}d_{ia} \right)^2
+i\lambda_i \left(\sum_a |d_{ia}|^2 -1 \right) \nonumber\\
&& - i \sum_{ab} N_{ab}^i \left(d_{ia}d_{ib} - P_{ab}^i \right) \Bigg \}
\nonumber\\
&& - \nu \sum_{<ij>}  \sum_{abcd} P_{ac}^i \Gamma^{abcd}_{ij}
P_{bd}^j \Bigg] \label{mfac2}
\end{eqnarray}
An integration over the auxiliary fields $N_{ab}^{i}$ shows that
$P_{ab}^i = \left<d_{ia} d_{ib}\right>$ and hence the nematic order
parameter $Q_{ab} = \sum_i \left(P_{ab}^i - \delta_{ab} \sum_c
P_{cc}^i/3\right)$ can be directly obtained in terms of the $P_{ab}^i$
fields.

Consequently, we can now seek the saddle point solution to the above
action.  At the saddle point, the constraint fields are
time-independent and the ${\bf d}_i$ fields can be integrated out.
Notice that in contrast to the clean system, the constraint fields
$\lambda_i$ are space-dependent. The mean-field action becomes
\begin{eqnarray}
S_{\rm MF} &=& \sum_i {\rm Tr}[\ln G_{MF}^{-1}] \nonumber\\
&& - \int d\tau \Bigg[ \sum_i
\Big (i \lambda_i  - i \sum_{ab} N_{ab}^i P_{ab}^i \Big )  \nonumber\\
&& + \nu \sum_{<ij>}
\sum_{abcd} P_{ac}^i \Gamma^{abcd}_{ij} P_{bd}^j \Bigg]    \label{mfac3} \\
G^{-1}_{MF}(\tau,\lambda_i,N^i) &=& \left[\left(-\partial_{\tau}^2
+ i\lambda_i \right)\delta_{ab} - i N_{ab}^i \right]  \label{mgf}
\end{eqnarray}
The saddle point  equations can now be obtained from Eq.\ \ref{mfac3},
\begin{eqnarray}
1 &=& \int \frac{d\omega}{2\pi} \sum_{a} \left[G_{aa}^{MF}
\left(\omega,\lambda_i, N^i\right) \right]  \label{mfe1} \\
 i N_{ab}^i &=& \nu \sum_{cd} \sum_{j} \Gamma_{acbd}^{ij} P_{cd}^i \label{mfe2}\\
P_{ab}^i &=&  \int \frac{d\omega}{2\pi} G^{MF}_{ab} \left(\omega,\lambda_i, N^i\right)
\label{mfe3}
\end{eqnarray}
and are solved numerically to obtain the mean field order parameter
$Q_{ab} = \sum_i \left(P_{ab}^i - \delta_{ab} \sum_c P_{cc}^i/3\right)$ for each
disorder realization.

\begin{figure}
\rotatebox{270}{
\includegraphics[width=6cm]{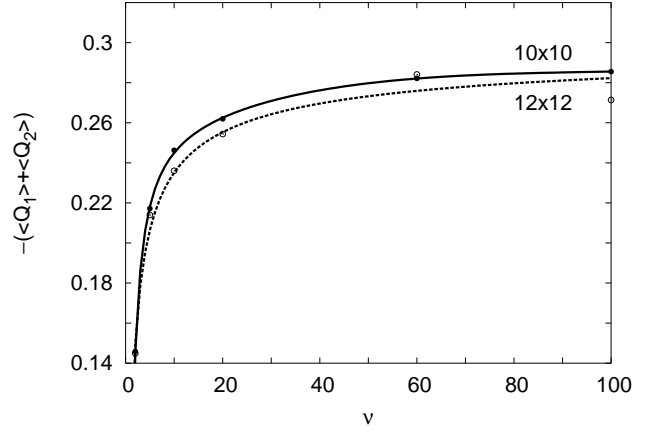}}
\caption{$-({\bar Q}_1+{\bar Q}_2)$ as a function of $\nu=g/4U$ for  
$({\bar \Lambda}^x-{\bar \Lambda}^z)/\delta\Lambda= 0.03$. This result,
obtained for  $ N_S=10\times 10$ and $N_S=12 \times 12$ lattices, shows
that for a wide range of $\nu > \nu_c $ the ground-state remains
biaxial nematic. [Note: ${\bar Q}_1 = \langle Q_1 \rangle$, 
${\bar Q}_2 = \langle Q_2 \rangle$]} \label{fig2}
\end{figure}

For a given disorder realization, we solve the set of 
mean-field equations Eqs.\ \ref{mfe1}, \ref{mfe2}, and \ref{mfe3} 
in order to find, for each lattice site
$i$, the fields $P_{ab}^{i}$ (with $a,b=\{x,y,z\}$). We
then average over $100$ disorder realizations
and for finite sizes $N_S= 10 \times 10$ and $12 \times 12$. We find
that the real valued solutions of the order parameter are diagonal
and obtain  $ {\bar Q}_a =\left<
Q_{a}\right>_{\rm disorder} \equiv (-({\bar Q}_1+{\bar Q}_2), 
{\bar Q}_1,{\bar Q}_2)$. The
result is shown in Fig.\ \ref{fig2} for $({\bar \Lambda}^x-{\bar
\Lambda}^z)/\delta \Lambda= 0.03$. We find, from the plot of $-({\bar Q}_1
+{\bar Q}_2)$ as a function of $\nu$, that the biaxial phase persists for a
wide range of $\nu > \nu_c \equiv 1$. This shows that the effect of
quantum fluctuations, at least at a saddle point level, does not
destabilize the biaxial phase as long as $\nu > \nu_c$. This
qualitative feature seems to be independent of system size, as can
be seen from Fig.\ \ref{fig2} and we expect it to hold in the
$N_S \rightarrow \infty$ limit.

\section{Ultracold spin-one atoms \label{atoms}}

In this section we first show that the low energy effective
Hamiltonian of spin-one ultracold atoms in the Mott insulating phase
can be mapped onto the rotor Hamiltonian (Eq.\ \ref{model1}) which
we analyzed earlier. Then we propose an imaging experiment on
ultracold atoms which can detect the biaxial nematic phase.

\subsection{Effective rotor model} \label{effective}

We consider a system of bosonic $S=1$ atoms in a disordered optical
lattice with spin-dependent confining potentials and
antiferromagnetic interaction between atoms. The many-body
Hamiltonian for this system is given in second-quantized notation by
\begin{eqnarray}
{\mathcal H} &=& \int d{\bf r}~\hat{\psi}^{\dagger}_{a}({\bf r})
\left(-\frac{\hbar^2}{2m}\nabla^2
+V_{a}({\bf r}) \right) \hat{\psi}_{a}({\bf r}) \nonumber \\
&&+\frac{1}{2} \int d{\bf r}d{\bf r'} \hat{\psi}^{\dagger}_{a}({\bf
r}) \hat{\psi}^{\dagger}_{a'}({\bf r'}) W({\bf r}-{\bf r'})
\hat{\psi}_{b}({\bf r'}) \hat{\psi}_{b'}({\bf r}),\nonumber \\
\label{mbH}
\end{eqnarray}
where $\hat{\psi}^{\dagger}_{a}({\bf r})$ is the boson field
operator that creates a particle with spin projection $a=\{-1,0,1\}$ at position
${\bf r}$, $V_{a}({\bf r})$ is the spin-dependent and spatially
disordered external potential, and $W({\bf r}-{\bf r'})= \delta({\bf
r}-{\bf r'})(U_0+U_2 {\bf S}_1 \cdot {\bf S}_2)$ is the two-body
interatomic potential \cite{pethick}. Here $U_{2} =
\frac{4\pi\hbar^2}{3m} (a_2 - a_0)$ and $U_{0} =
\frac{4\pi\hbar^2}{3m} (2a_2 + a_0)$ are the on-site interactions,
$a_{2(0)}$ are the s -wave scattering lengths in the $S=2(0)$ spin
channels, and $m$ is the mass of the atoms.
For example, for $^{23}$Na, the scattering lengths are 
$a_2=(52 \pm 5)a_B$ and $a_0=(46 \pm 5)a_B$ where $a_B$ is the 
Bohr radius\cite{Adi1} so that $U_2/U_0 \sim 0.04$.
In what follows, we
shall be interested in the Mott states of the spin-one bosons which
occur in the limit of deep lattice potential with $V_1 = V_{-1} \ne
V_0$.

A spin-dependent disordered potential $V_{a}({\bf r})$ can be
generated by superposing a speckled laser field on a sinusoidal
spin-dependent lattice potential $V_{0a}({\bf r})$. The spin
dependence of the sinusoidal potential can be achieved by tuning the
laser frequency close to the hyperfine splitting (but far away from
the fine structure splitting) of the atoms
\cite{jaksch2,brennen,liu,mandel2,carusotto,duradev}. In appendix \ref{app},
we propose a method to obtain a spin-dependent lattice potential
with $V_1 = V_{-1} \ne V_0$. Generation of the
disordered potential $\delta V_{a}({\bf r})$ can be achieved by
reflecting a laser at the same frequency but with a much lower intensity
off a speckled mirror. The net lattice potential seen by the atoms
at spin state $a$ is then $V_{a}({\bf r}) =V_{0a}({\bf r})+
\delta V_{a}({\bf r})$. In what follows, we shall consider the
situation where the speckled field laser is weak compared to the one
generating the sinusoidal potential such that $\delta V_{a} \ll
V_{0a}$.

For free ultracold atoms in an optical lattice, the energy
eigenstates are Bloch wave functions and a superposition of these
Bloch states yields a set of Wannier functions which are well
localized on the individual lattice sites for deep lattices
\cite{greiner}. The energies involved in the system dynamics being
small compared to excitation energies to the second band, we expand
the boson field operators in the Wannier basis and keep only the
lowest states, $\psi_{a}({\bf r})=\sum_i b_{ia} w({\bf r}-{\bf
r}_i)$. Consequently, the many-body Hamiltonian (\ref{mbH}) reduces
to \cite{Demler1}
\begin{eqnarray}
{\mathcal H} &=& \frac{U_0}{2} \sum_{i} {\hat n}_{i} ({\hat
n}_{i}-1) + \frac{U_2}{2} \sum_{i} \left({\bf S}_i^2 - 2
\hat{n}_i\right)
-\mu \sum_{i} \hat{n}_i \nonumber\\
&& -\sum_{\langle ij \rangle}\sum_{a} \left( b_{ia}^{\dagger}
{\tilde t}^{ij}_{a} b_{ja} + { \rm h.c} \right), \label{ham1}
\end{eqnarray}
where $b_{ia}$ is the spin-one boson operator at site $i$ with 
spin projection
$a =\{-1,0,1\}$, ${\hat n}_i = \sum_{a} b_{ia}^{\dagger} b_{ia}$ is
the boson density at site $i$, ${\hat {\bf S}}_i = \sum_{a b }
b_{ia}^{\dagger} {\mathcal S}_{ab} b_{i b}$ is the spin operator
(${\mathcal S}$ being the spin rotation matrices for spin-one bosons),
$\mu$ is the chemical potential, and ${\tilde t}_{a}^{ij}$ are given
by
\begin{equation}
{\tilde t}_{a}^{ij} = \int d{\bf r} w^{*}({\bf r}-{\bf r}_i)
\left(-\frac{\hbar^2}{2m}\nabla^2 +V_{a}({\bf r}) \right) w({\bf
r}-{\bf r}_j).
\end{equation}
so that  $\tilde{t}_1 = \tilde{t}_{-1} \ne \tilde{t}_0$ for $V_1 = V_{-1} \ne V_0$.

Note that for weak speckled fields, we always have $\sigma_{t}
/{\bar t}_a \ll 1$, where $\sigma_t$ and ${\bar t}_a$ are the
standard deviation and average of the hopping coefficient $\tilde{t}_a$.
However, in this setup, one can always tune the sinusoidal potential
$V_{0a}$ so that we are in a regime where $|{\bar t}_1-{\bar t}_0|
/\sigma_{t} \ll 1 $. As we shall see, this is precisely the
regime where one expects to see the biaxial phase. Also we note that
due to the on-site disordered potential, the chemical potential
$\mu_i$ should also be site-dependent. However, $\mu_i$ has only a
power law dependence on disorder by comparison to an exponential
dependence for the hopping coefficients. Consequently, the standard
deviation of $\mu_i$ is small and does not influence the nature of
the Mott states. So we replaced $\mu_i$ by its average value $\mu$ in
Eq.\ \ref{ham1}. Notice that this approximation is valid only when
the potential due to the speckled field is weak compared to the one due
to the sinusoidal field.

Next, following Ref.\ \onlinecite{Adi1}, we switch to a
representation where the boson operators transform as vectors under
spin rotation
\begin{eqnarray}
b_{iz} &=& b_{i0}, \, \, b_{ix} =\frac{1}{\sqrt{2}}(b_{i\,
-1}-b_{i\,1}),
\, \, b_{iy} = \frac{-i}{\sqrt{2}} (b_{i\,-1}+b_{i1}).\nonumber\\
\end{eqnarray}
Using these operators, the kinetic term in (\ref{ham1}) is rewritten
as
\begin{equation}
-\sum_{\langle ij \rangle}\sum_{a \in \{x,y,z\}} \left(
b_{ia}^{\dagger} t^{ij}_{a} b_{ja} + { \rm h.c} \right),
\end{equation}
with $t_x^{ij} = t_y^{ij} = 2{\tilde t}_{1}^{ij}$ and 
$t_z^{ij} = {\tilde t}_{0}^{ij}$.

We now consider this spin-one Bose-Hubbard model in the limit where
$N \gg 1$ and we are in the Mott phase of the bosons with a large
odd number ($N$) bosons per site. A straightforward generalization
of the analysis of Ref.\ \onlinecite{Adi1}, shows that the low
energy effective Hamiltonian in this limit can be mapped on to a
rotor model. Using the decomposition $ b_{ia} = d_{ia} a_i$, where
the boson operator $a_i$ changes the number of particles $N_i$ but
not the orientation of the boson spin given by ${\bf d}_i$, we
obtain in second order perturbation theory \cite{Adi1,Demler1}
\begin{eqnarray}
{\mathcal H}_{\rm eff} &=& \frac{U_2}{2} \sum_i S_i^2 -\frac{2 N^2
{\bar t}_x^2}{U_0} \sum_{<ij>} \left( \sum_{a=x,y,z} d_{ia}
\frac{t_a^{ij}}{{\bar t}_x} d_{ja} \right)^2. \nonumber\\
\label{ham2}
\end{eqnarray}
Eq.\ \ref{ham2}
has to be supplemented with the constraint $N_i + S_i = {\rm even}$ where
$N_i$ is the number of bosons at site $i$, and 
$S_i= i \epsilon_{ijk} d_j \frac{\partial}{\partial d_k}$ is the total
spin which can be identified as the rotor angular momentum. 
However, the constraint $N_i + S_i = {\rm even}$ becomes 
irrelevant in the small $U_2$ limit \cite{Adi1}.
Therefore in this limit, the effective low energy Hamiltonian (Eq.\
\ref{ham2}) can be directly mapped to the rotor model (Eq.\
\ref{model1}) with the identification $U_2/2 \rightarrow U$, $2 N^2
{\bar t}_x^2/U_0 \rightarrow g$ and $t_a^{ij}/{\bar t}_x \rightarrow
\Lambda_a^{ij}$. For the biaxial nematic phase to occur, we
therefore need a window where the conditions $4 N^2{\bar t}_x^2/U_2
U_0 \gg 1$, and $N{\bar t}_x/U_0 < 1$ are simultaneously satisfied.
The second condition arises here since we need to be away from the
superfluid transition point of boson systems.
These conditions can be expected to be easily satisfied.
For example, in a deep lattice ($V=10 E_R$ where $E_R$ is the recoil energy) 
set by red detuned light ($\lambda = 985~$nm) and containing about $10$ 
sodium atoms per well, $4 N^2{\bar t}_x^2/U_2 U_0 \sim 14$, 
and $N{\bar t}_x/U_0 \sim 0.4$.

\subsection{Detection of biaxial nematic order \label{detection}}

In this section, we propose a method to detect experimentally
biaxial nematic order in a condensate of spin-one cold atoms.
Consider the atoms being in a Mott state which has biaxial nematic
order parameter. First, as is customary in most of the experiments
in ultracold atoms \cite{greiner}, we switch off both the lattice
potential and the trap, and let the atoms expand freely. We then
probe the expanding cloud by a right-circular ($\sigma_+$) 
polarized laser beam. The
dielectric tensor of the atoms as seen by the laser beam is given by
\cite{carusotto}
\begin{equation}
\langle\epsilon_{jk}\rangle=\delta_{jk}+c_0\langle\rho\rangle\delta_{jk}
-ic_1\varepsilon_{jkl}\langle S_l\rangle+c_2\langle Q_{jk}\rangle,
\end{equation}
where the coefficients $c_{a=\{0,1,2\}}$ depends on the laser
frequency and $\langle\rho\rangle$, $\langle{\bf S}\rangle$ and
$\langle Q_{jk}\rangle$ are the density,
average spin and the nematic order parameters of the atomic cloud.
If the laser frequency is too far detuned from hyperfine splitting
frequency of the atomic levels, $c_2$ vanishes and the nematic order
is not probed. On the other hand, if the laser frequency is not
detuned enough from the hyperfine splitting frequency, there will be
significant absorption which will weaken the intensity of the
transmitted light. As shown in Ref.\ \onlinecite{carusotto}, there
indeed exist a window for several spin-one atom species where the
imaging can be done.

If the expanding cloud is sufficiently optically thin and
homogeneous, the polarization of the transmitted beam (taken to be
propagating along the $z$ axis) is
\begin{equation}
{\bf \wp}_{\rm out} = e^{\frac{i\omega\Delta}{c}} [{\bf
1}+\frac{i\omega}{c}\int_{0}^{\Delta} dz~(\sqrt{\epsilon_{(xy)}}
-{\bf 1})]~{\bf \wp}_{\rm in},
\end{equation}
where ${\bf \wp}_{in}=(\wp_x,\wp_y)$ is the two-component
polarization vector of the incoming laser beam, $\epsilon_{(xy)}$ is
the reduced dielectric tensor in the $(xy)$ plane and $\Delta$ is
the thickness of the medium. As discussed in Ref.\ \onlinecite{carusotto},
the presence of spin order $\langle {\bf S} \rangle$ in the atom cloud gives a phase
shift to the atoms whereas a nematic order leads to a left-circular ($\sigma_-$)
polarized component in the transmitted beam. So, if we shine on the 
spin-one sample a beam of pure $\sigma_+$ light in such a way
that the principal axes $Q_1$ and $Q_2$ of the nematicity ellipsoid
are orthogonal to the direction of the propagating beam, the intensity 
of the $\sigma_-$ component of the transmitted beam is given by
\begin{equation}
I_-=|~\alpha_+~\frac{i\omega c_2}{4c} \int_0^{\Delta}
dz~(\langle Q_1 \rangle-\langle Q_2 \rangle)~|^2
\end{equation}
where $\alpha_+$ is the amplitude of the incoming beam. Note that
this method distinguishes between uniaxial and biaxial nematic
ground states. In the uniaxial state $\langle Q_1\rangle=\langle
Q_2\rangle$ and $I_-=0$; however, for a biaxial nematic ground state
$\langle Q_1\rangle\neq\langle Q_2\rangle$ so $I_-\neq 0$. Thus
passing the transmitted beam through a crossed polarizer, one should
be able to measure $I_-$ and hence detect the presence of a biaxial
nematic state.

\section{Conclusion \label{conclu}}

We have studied a disordered O(2) rotor model with quadrupolar
interaction and demonstrated that the model exhibits a biaxial
nematic phase in the disordered average sense. It is demonstrated
that within mean-field analysis, the biaxial nematic phase is stable
against small quantum fluctuations. Such models are shown to be
realized in the Mott phase of spin-one ultracold bosons in optical lattices 
with spin-dependent disordered potential 
in the limit of large number of bosons per site. We have
also suggested an experiment which can, using laser imaging of the
spin-one atoms, detect the biaxial nematic phase.

\begin{acknowledgments}
This work was supported by NSERC, the Canadian Institute for Advanced
Research, the Canada Research Chair Program (YBK, KS, JSB), and  
Le Fonds qu\'eb\'ecois de la recherche sur la nature et les technologies (JSB). 
YBK thanks Chetan Nayak for discussions that sparked his interest in biaxial nematic
phases, KS thanks Duncan O'dell and Ying-Jer Kao for helpful discussions.

\end{acknowledgments}

\appendix

\section{Spin-dependent optical lattice \label{app}}

We propose here a method to create a spin-dependent optical 
square lattice. Using our approach, trapped bosons 
with $S_z=\{-1,1\}$ experience the same potential, $V_{-1}=V_{1}$, 
while bosons with $S_z=0$ are subject to a 
different potential $V_0$. 

Consider atoms with total angular momentum $S=1$
interacting with a configuration of laser beams producing an electric 
field ${\bf E}({\bf r})$. Building on previous work
for $S=1/2$ particles\cite{duradev}, we deduce that atoms with 
$S=1$ experience an external potential of the form
\begin{equation}
V_{\alpha\beta}({\bf r})=V({\bf r})\delta_{\alpha\beta}
+{\bf B}({\bf r})\cdot\hat{\bf S}_{\alpha\beta}
+N_{ij}({\bf r})\hat{G}_{ij\alpha\beta},
\label{extpot}
\end{equation}
with $\alpha,\beta=\{-1,0,1\}$. In Eq.\ \ref{extpot}, the scalar 
potential $V({\bf r})$ is proportional to the light
intensity, the vector field ${\bf B}({\bf r})$ is proportional 
to the electromagnetic spin\cite{duradev} and couples to the total atomic angular 
momentum operator $\hat{{\bf S}}$, and the second-rank tensor $N_{ij}({\bf r})$ 
is proportional to the light nematicity\cite{carusotto} and couples to the quadrupole moment
operator $\hat{G}_{ij}$:
\begin{eqnarray}
V({\bf r}) &=& b_0 {\bf E}^*({\bf r})\cdot{\bf E}({\bf r}) \nonumber \\
{\bf B}({\bf r}) &=& -i b_1 {\bf E}^*({\bf r})\times{\bf E}({\bf r})\nonumber \\
N_{ij}({\bf r}) &=& b_2\big[\frac{1}{2}(E^*_i({\bf r})E_j({\bf r})
+E^*_j({\bf r})E_i({\bf r})) \nonumber \\
&& ~~~~~ -\frac{1}{3}{\bf E}^*({\bf r})\cdot{\bf E}({\bf r}) \delta_{ij}\big] \nonumber \\
\hat{\bf S}_{\alpha\beta} &=& \langle 1,\alpha|\hat{\bf S}|1,\beta\rangle \nonumber \\
\hat{G}_{ij\alpha\beta} &=& \langle 1,\alpha|\left[\frac{1}{2}(\hat{S}^{\dagger}_i\hat{S}_j
+\hat{S}^{\dagger}_j\hat{S}_i)
-\frac{1}{3}\hat{\bf S}^2 \delta_{ij}\right]|1,\beta\rangle. \nonumber \\
\end{eqnarray}
The coefficients $b_{0,1,2}$ are functions of the
light frequency and the atomic structure. To obtain an 
effective coupling to the light nematicity (i.e. a large enough $b_2$ value), 
one needs to tune the laser frequency close to the hyperfine splitting 
of the atoms, but far from the fine structure splitting such that 
$b_1 \ll b_2, b_0$.

Then, to generate the optical square lattice (say in the $xy$ plane), we use two orthogonal 
pairs of counter propagating monochromatic lasers, and choose these equal intensity light 
fields to be linearly polarized in the $z$ direction. The total electric field 
produced by this configuration is thus given by
\begin{eqnarray}
{\bf E}(t,x,y)= 2~E_0~\hat{\bf z}~e^{i\omega t} 
\left[e^{i\phi_x} \cos(kx) + e^{i\phi_y} \cos(ky)\right],\nonumber \\
\label{fieldconf}
\end{eqnarray}
where $k$ is the wavevector, and $\phi_x$, $\phi_y$ are the initial phases
for the electric field propagating in the $x$ and $y$ directions respectively. 
We choose the difference between these two initial 
phases $\Delta \phi = \phi_x-\phi_y$ to 
be equal to  $\pi/2$. Using this electric field configuration, we find the
electromagnetic spin to be zero and the 
external potential to be
\begin{eqnarray}
V_{\alpha\beta}(x,y) &=&
4 |E_0|^2 \left(\cos^2(kx)+\cos^2(ky)\right) \times \nonumber \\
&&~ \left( (b_0-\frac{2}{3} b_2) \delta_{\alpha\beta}
+ b_2 \langle 1,\alpha|\hat{S}_z^2|1,\beta\rangle \right). \nonumber \\
\end{eqnarray}
Hence, only the diagonal terms of the external potential tensor are non-zero and are given
by
\begin{eqnarray}
V_{00} &=& A(x,y) (b_0-\frac{2}{3} b_2) \nonumber \\
V_{11} &=& A(x,y) (b_0+\frac{1}{3} b_2) \nonumber \\
V_{-1-1} &=& A(x,y) (b_0+\frac{1}{3} b_2),
\end{eqnarray}
where $A(x,y)= 4 |E_0|^2 (\cos^2(kx)+\cos^2(ky))$. As a result, 
we obtain a spin-dependent optical square 
lattice with  $V_{1}=V_{-1}\neq V_{0}$.

%case, since we need atoms in each spin states to see an separate
%disorder potential. This can be achieved by superposing two speckled
%laser fields. The first laser is to be tuned at frequency far away
%from the fine or hyperfine structure splitting of the atoms giving
%rise to a spin independent disordered potential $\delta V_1({\bf
%r})$ while the second is tuned at the same frequency as the laser
%used to generate the sinusoidal potential leading to a potential
%$\delta V_{2a}({\bf r})$ for the atoms \cite{comment3}. The net
%lattice potential seen by the atoms at spin state $a$ is then
%$V_{\sigma}({\bf r}) =V_{0a}({\bf r})+ \delta V_1({\bf r}) + \delta
%V_{2a}({\bf r})$. Thus in this setup the disorder potential for the
%atom at a given site is random number drawn from separate
%independent distributions for each spin state with approximately the
%same variance. In what follows, we shall consider the situation
%where the speckled field laser is weak compared to the one
%generating the sinusoidal potential such that $\delta V_1, \delta
%V_{2a} \ll V_{0a}$.


\begin{thebibliography} {99}


\bibitem{dg} See chapter 2 in P.G. de Gennes and J. Prost {\it The Physics
             of Liquid Crystals} (Oxford University Press, New York, 1993).

\bibitem{merkel} K. Merkel {\it et\,al.}, Phys. Rev. Lett. {\bf 93},
                 237801 (2004).

\bibitem{Mermin} For a review on defect structures of biaxial nematic see
                 N.D. Mermin, Rev. Mod. Phys. {\bf 51} 591 (1979).

\bibitem{halboth} C.J. Halboth and W. Metzner, Phys. Rev. Lett. {\bf 85}, 5162 (2000).

\bibitem{oganesyan} V. Oganesyan, S.A. Kivelson, and E. Fradkin, 
                    Phys. Rev. B {\bf 64}, 195109 (2001).

\bibitem{ybk} H.Y. Kee and Y.B. Kim, Phys. Rev. B {\bf 71}, 184402 (2005).

\bibitem{kivelson} S.A. Kivelson, I.P. Bindloss, E. Fradkin, V. Oganesyan, J.M. Tranquanda, 
A. Kapitulnik and C. Howald, Rev. Mod. Phys. {\bf 75}, 1201 (2003).

\bibitem{andreev} A.F. Andreev and I.A. Grishchuk, Sov. Phys. JETP {\bf 60}, 267 (1984).

\bibitem{chandra} P. Chandra and P. Coleman, Phys. Rev. Lett. {\bf 66}, 100 (1991).

\bibitem{gorkov} L.P. Gor'kov and A. Sokol, Phys. Rev. Lett. {\bf 69}, 2586 (1992).


\bibitem{Demler1} E. Demler and F. Zhou, Phys. Rev. Lett. {\bf 88},
                 163001 (2002).

\bibitem{Adi1}  A. Imambekov, M. Lukin, and E. Demler, Phys. Rev. A {\bf 68}, 063602 (2003).

\bibitem{snoek1}  F. Zhou and M. Snoek, Ann. Phys. {\bf 308}, 692 (2003).


\bibitem{horak} P. Horak, J.Y. Courtois, and G. Grynberg,
                Phys. Rev. A. {\bf 58}, 3953 (1998).

\bibitem{lye} J.E. Lye, L. Fallani, M. Modugno, D. Wiersma, C. Fort, and
              M. Inguscio, Phys. Rev. Lett. {\bf 95}, 070401 (2005).

\bibitem{comment2} In the case of $N_S= 20 \times 20$, we have studied four different
systems sizes starting from $20 \times 8$ up to $20 \times 14$, and
extrapolated the order paramater to its $20 \times 20$ value.

\bibitem{kleinert} H. Kleinert {\it Path Integrals in Quantum Mechanics, Statistics,
                   Polymer Physics and Financial Markets}
                   (World Scientific, Singapore, 2004).

\bibitem{pethick} C.J. Pethick and H. Smith {\it Bose-Einstein Condensation in Dilute Gases}
                  (Cambridge University Press, Cambridge, 2002).


\bibitem{jaksch2} D. Jaksch, H. Briegel, J. Cirac, C. Gardiner, and P. Zoller,
                  Phys. Rev. Lett. {\bf 82}, 1975 (1999).

\bibitem{brennen} G. Brennen, C. Caves, P. Jessen, and I. Deutsch,
                  Phys. Rev. Lett. {\bf 82}, 1060 (1999).

\bibitem{liu}     W.V. Liu, F. Wilczek, P. Zoller, Phys. Rev. A {\bf 70}, 033603 (2004).

\bibitem{mandel2} O. Mandel, M. Greiner, A. Widera, T. Rom, T. H\"ansch, and I. Bloch,
                  Phys. Rev. Lett. {\bf 91}, 010407-1 (2003).

\bibitem{carusotto} I. Carusotto and E. Mueller, J. Phys. B {\bf 37}, S115 (2004).

\bibitem{duradev} A. Duradev, R.B. Diener, I. Carusotto, and Q. Niu, Phys. 
                  Rev. Lett. {\bf 92}, 153005-1 (2004).


%\bibitem{comment3}  If the two lasers are sufficiently detuned from
%each other, the cross terms in the potentials arising from the
%superposition of their fields vanish within rotating wave
%approximation.

\bibitem{greiner} M. Greiner, O. Mandel, T. Esslinger, T.W. H\"ansch, and I. Bloch,
                  Nature {\bf 415}, 39 (2002); D. Jaksch, C. Bruder, J.I. Cirac, C.W. Gardiner,
                  and P. Zoller, Phys. Rev. Lett. {\bf 81}, 3108 (1998).


\end{thebibliography}
\end{document}